\documentclass[aps,prapplied,twocolumn,superscriptaddress,floatfix]{revtex4-1}

\usepackage{graphicx}
\usepackage{amsmath}
\usepackage{bm}
\usepackage{color,soul}

\definecolor{mygreen}{rgb}{0.0, 0.7, 0.2}

\begin{document}
\setlength{\textfloatsep}{12pt}
%\preprint{}

\title{Conversion efficiency in Kerr microresonator optical parametric oscillators: From three modes to many modes}


\author{Jordan R. Stone}
\email{jstone12@umd.edu}
\affiliation{Joint Quantum Institute, NIST/University of Maryland, College Park, MD 20742}
\affiliation{National Institute for Standards and Technology, Gaithersburg, MD 20899}

\author{Gregory Moille}
\affiliation{Joint Quantum Institute, NIST/University of Maryland, College Park, MD 20742}
\affiliation{National Institute for Standards and Technology, Gaithersburg, MD 20899}

\author{Xiyuan Lu}
\affiliation{National Institute for Standards and Technology, Gaithersburg, MD 20899}
\affiliation{Institute for Research in Electronics and Applied Physics and Maryland NanoCenter, University of Maryland,
College Park, MD 20742, USA}

\author{Kartik Srinivasan}
\affiliation{Joint Quantum Institute, NIST/University of Maryland, College Park, MD 20742}
\affiliation{National Institute for Standards and Technology, Gaithersburg, MD 20899}

%\collaboration{Xu Yi}

\date{\today}

\begin{abstract}
We study optical parametric oscillations in Kerr-nonlinear microresonators, revealing an intricate solution space -- parameterized by the pump-to-signal conversion efficiency -- that arises from an interplay of nonlinear processes. Using a three-mode approximation, we derive an efficiency-maximizing relation between pump power and frequency mismatch. To move beyond a three-mode approximation, a necessity for geometries such as integrated microring resonators, we numerically simulate the Lugiato-Lefever Equation that accounts for the full spectrum of nonlinearly-coupled resonator modes. We observe and characterize two nonlinear phenomena linked to parametric oscillations in multi-mode resonators: Mode competition and cross phase modulation-induced modulation instability. Both processes may impact conversion efficiency. Finally, we show how to increase the conversion efficiency by tuning the microresonator loss rates. Our analysis will guide microresonator designs that aim for high conversion efficiency and output power. %Our analysis provides guidance on how the resonator dispersion and balance of intrinsic and coupling losses should be engineered to realize high conversion efficiency at a targeted pump power. 
\end{abstract}

%\pacs{}
%\keywords{}

\maketitle

%\section{}

\section{Introduction}
Integrated photonics offers scalable options for generating, processing, and routing optical signals within classical and quantum networks \cite{liang2010recent,agrell2016roadmap,sipahigil2016integrated,elshaari2020hybrid,jin2021hertz}. In general, optical processors apply linear and/or nonlinear operations to light. A notable case is the optical microresonator, whose small size and large quality factor ($Q$) work to intensify circulating light and promote efficient nonlinear interactions \cite{vahala2003optical,yang2018bridging}. Indeed, microresonators host a veritable zoo of nonlinear eigenstates, including soliton frequency combs \cite{kippenberg2018dissipative}, Raman frequency combs \cite{liu2018integrated}, Hz-linewidth lasers based on stimulated Brilluoin scattering \cite{loh2015dual}, $\chi^{(2)}$ and $\chi^{(3)}$-type parametric oscillators \cite{bruch2019chip, lu2019milliwatt}, and more for applications in communications, timekeeping, and sensing \cite{spencer2018optical,marin2017microresonator,lai2020earth,newman2019architecture}. 

Many of the experiments cited above were motivated by a high demand for coherent light sources on a chip. One important type of coherent source is the optical parametric oscillator (OPO), which is often employed to reach wavelengths not directly accessible by conventional laser gain \cite{vodopyanov2000zngep,mieth2014tunable}. Optical parametric oscillations occur in $\chi^{(3)}$-nonlinear media when vacuum fluctuations are amplified by stimulated four wave mixing (FWM), if the FWM gain exceeds the resonator losses \cite{boyd2020nonlinear}. Degenerately-pumped OPOs are a special case in which two frequency-degenerate pump photons are converted into one higher-frequency signal photon and one lower-frequency idler photon. In principle, a degenerately-pumped OPO can generate coherent light within the frequency range $\textrm{DC}-2\omega_{\rm{p}}$, where $\omega_{\rm{p}}$ is the pump laser frequency \cite{lu2021considering}. Hence, a chip-scale, degenerately-pumped OPO could offer superior scalability, higher efficiencies, and broader spectral coverage than alternatives. It would be readily implemented in miniaturized technologies, from optical clocks in which the OPO could be tuned to clock-type or cooling-type atomic transitions, to quantum processors in which the OPO could be tuned to qubit frequencies. 

\begin{figure*}
    \centering
    \includegraphics[width=500 pt]{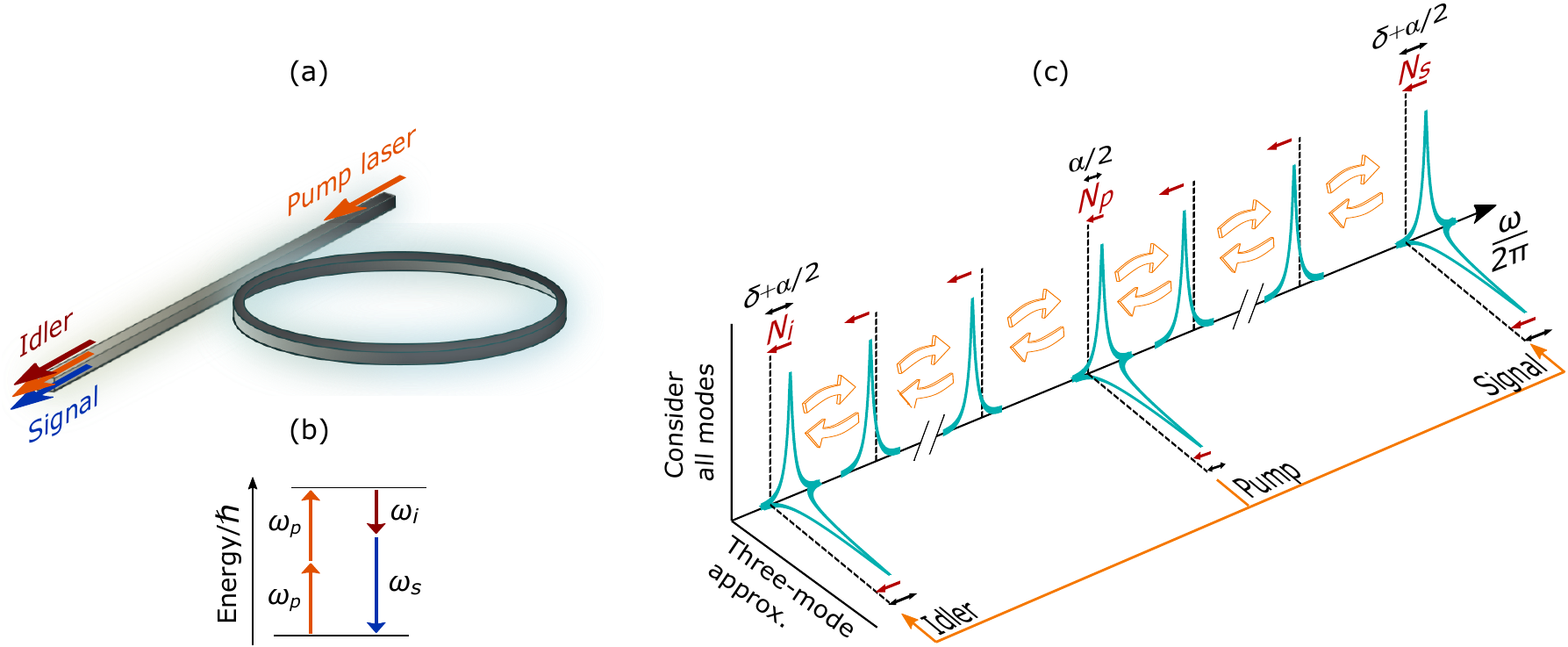}
    \caption{Introduction to the microresonator-based, degenerately-pumped optical parametric oscillator ($\mu$OPO). (a) Schematic of a microring resonator coupled to an access waveguide that carries the input and output fields. (b) Energy diagram for the degenerate four wave mixing (FWM) process that drives parametric oscillation. (c) Depictions of mode spectra, nonlinear couplings, and frequency shifts in both a three-mode approximation (TMA) and multi-moded model. Dashed lines correspond to zero frequency mismatch. Red arrows indicate mode frequency shifts induced by Kerr nonlinearity. The TMA considers FWM between only the pump, signal, and idler modes, while multi-moded models account for the nonlinear couplings (orange, hollow arrows) between all mode sets conducive to FWM.} 
    \label{fig:intro}
\end{figure*}

Recently, several experiments have been reported that advance the microresonator-based, degenerately-pumped OPO ($\mu$OPO) and make real headway towards a chip-scale, wavelength-by-design light source. Achievements include sub-milliwatt oscillation thresholds \cite{lu2019milliwatt}, octave-spanning and tunable spectra \cite{sayson2019octave,tang2020widely}, visible-light generation spanning red to green \cite{lu2020chip, domeneguetti2021parametric}, and a $\mu$OPO that uses a 2D photonic crystal cavity \cite{marty2021photonic}.  Nonetheless, the reported pump-to-signal conversion efficiencies are typically (except in instances involving narrow spectral bandwidths) $<0.1$\% -- a nonstarter for applications \cite{lu2020chip, sayson2019octave, domeneguetti2021parametric, tang2020widely}. Indeed, while the demand for efficiency calls for a deeper understanding of the underlying nonlinear physics, experiments have so far relied on a simplified theoretical framework for $\mu$OPOs. For instance, frequency matching is considered in either a cold-cavity limit \cite{lu2019milliwatt}, or otherwise only accounts for a populated pump mode \cite{sayson2019octave}. A more accurate description of frequency matching should account for the exact distribution of intraresonator photons. Moreover, analyses have relied on a three-mode approximation (TMA), in which only the pump, signal, and idler modes interact through Kerr nonlinearity  \cite{sayson2019octave,hansson2013dynamics}. Of course, real microresonators comprise a more complex spectrum of modes that are nonlinearly coupled together. As a result, there is presently a gap between theoretical and experimental progress revolving around $\mu$OPOs; ultimately, there is little theoretical basis on which to design microresonators to meet end-user demands. 

Here, we construct a generalized $\mu$OPO solution space; thereby, we reveal connections between the $\mu$OPO state and experimental parameters, and we identify processes that limit conversion efficiency. We adopt a model based on the Lugiato-Lefever Equation (LLE) \cite{coen2013modeling,chembo2013spatiotemporal} and support our main numerical results with theoretical analyses. In the next section, we explain our modeling and present simulation results using a TMA. Then, we expand the model to include a spectrum of nonlinearly-coupled resonator modes. We demonstrate two nonlinear phenomena that cannot be explained within a TMA. In the first, a mode competition takes place between multiple signal and idler mode pairs. In the second, modulation instability induced by cross-phase modulation constrains the $\mu$OPO conversion efficiency. Finally, we propose two strategies for increasing the $\mu$OPO conversion efficiency and output power. Surprisingly, when comparing microresonators with different loss rates but identical geometries, we find that the resonator with greater losses will, in some cases, promote higher efficiency.  %Our results constitute an initial theoretical foundation for $\mu$OPO, and they will guide designs for efficient and high-power $\mu$OPO systems that are crucial for applications.  

\section{Modeling the $\mu$OPO: The Lugiato-Lefever Equation, three-mode approximation, and dispersion}

To study $\mu$OPOs, we consider a microring resonator coupled to an optical access waveguide and pumped by a continuous-wave (CW) laser, as depicted in Fig. \ref{fig:intro}a. This structure supports whispering gallery modes, in which azimuthal modes are grouped into families sharing a transverse spatial mode profile. Modes within a family are spaced (in the frequency domain) by a free-spectral range (FSR) that is inversely proportional to the ring circumference, $L$. In our model, we consider a single mode family and denote its resonant frequencies as $\omega_{\rm{\mu}}$, where $\mu$ is the azimuthal mode number shifted to make $\omega_{\rm{0}}$ the frequency of the pumped mode. The pump laser has frequency $\omega_{\rm{p}}$ and waveguide power $P_{\rm{in}}$, and the intraresonator field, $a$, obeys the Lugiato-Lefever Equation (LLE) \cite{chembo2013spatiotemporal}: 
\begin{equation}{\label{eq:LLE}}
    \frac{da}{dt}=\sqrt{\frac{\kappa_{\rm{c}}(0)}{\hbar \omega_{\rm{p}}}P_{\rm{in}}}-\left(\frac{\kappa_{i}}{2}+i\frac{\kappa(0)}{2}\alpha-ig_{0}|a|^2\right)a-i\mathcal{D}(\mu)\tilde{a},
\end{equation}
where $|a|^2$ gives the intraresonator energy in units of photon number, $\kappa_{\rm{c}}(\mu)$ is the mode-dependent coupling rate to the access waveguide, $\kappa_{\rm{i}}$ is the mode-independent intrinsic loss rate, $\kappa(\mu)=\kappa_{\rm{i}}+\kappa_{\rm{c}}(\mu)$ is the mode-dependent total loss rate, $\alpha=\frac{\omega_{\rm{0}}-\omega_{\rm{p}}}{\kappa(0)/2}$ is the normalized pump-resonator frequency detuning, $g_{\rm{0}}=\frac{n_{\rm{2}}c\hbar\omega_{\rm{0}}^2}{n^2V}$ is the nonlinear gain per photon, $n_{\rm{2}}$ is the Kerr index, $c$ is the speed of light in vacuum, $n$ is the refractive index, $V$ is the mode volume, and $\mathcal{D}(\mu)=\omega_{\rm{\mu}}-(\omega_{\rm{0}}+\mu D_{\rm{1}})+i\kappa_{\rm{c}}(\mu)$, where $D_{\rm{1}}=2\pi\times \textrm{FSR}$; $\tilde{a}$ indicates that operations to $a$ are performed in the frequency domain. Notably, the integrated dispersion, $D_{\rm{int}}$, is contained in Eq. \ref{eq:LLE} as $D_{\rm{int}}=Re(\mathcal{D})/\kappa(0)$. For concreteness, we use $n_{\rm{2}}=2.4 \times 10^{-19}$ m$^2$/W and $n=1.9$, which are typical values for silicon nitride (SiN) microrings, and $\omega_{\rm{p}}\approx 2\pi \times 384$ THz ($\approx$ 780 nm wavelength). Unless otherwise stated, we use $\kappa_{\rm{i}}=2\pi \times 200$ MHz and $\kappa_{\rm{c}}=\kappa_{\rm{i}}$ (i.e., critical coupling).

\begin{figure*}
    \centering
    \includegraphics[width=500 pt]{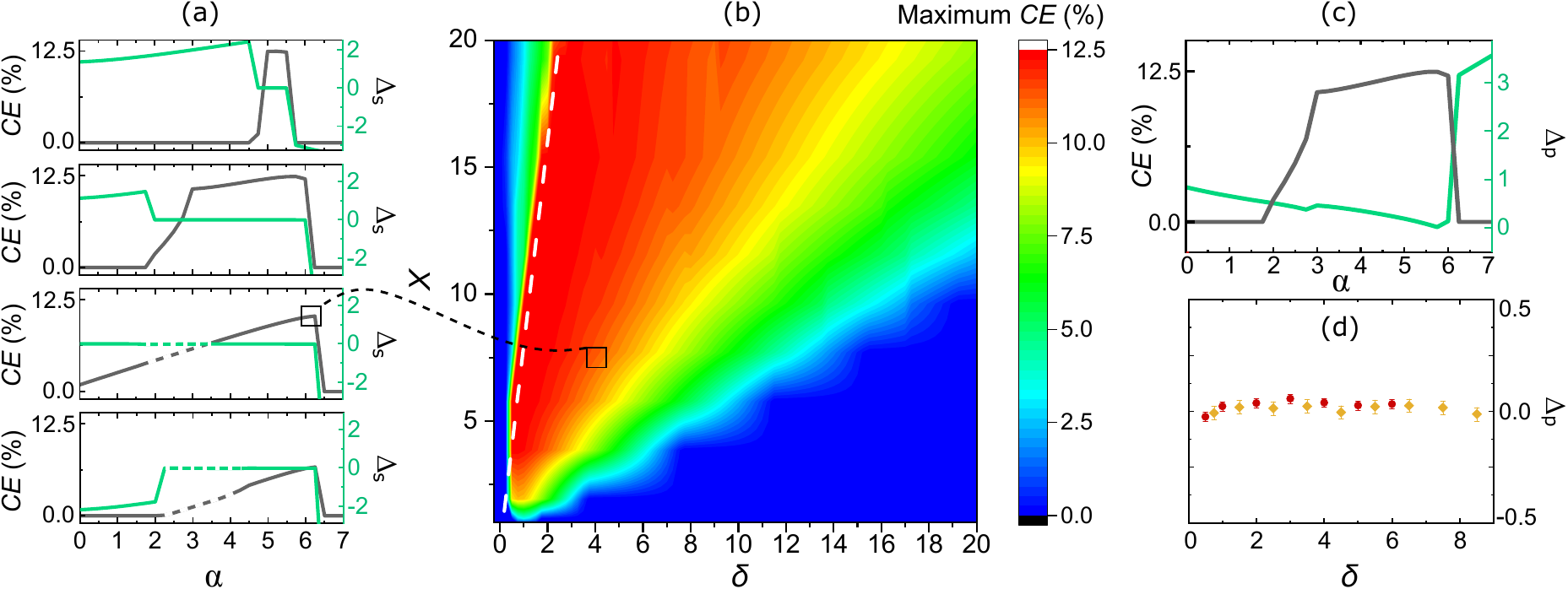}
    \caption{Study of the $\mu$OPO conversion efficiency using a three-mode approximation. (a) Conversion efficiency, $CE$ (gray), and effective mismatch, $\Delta_{s}$ (green), versus the pump-resonator frequency detuning, $\alpha$, for a pump laser power $P_{\rm{in}}=20~\rm{mW}$. From top to bottom, the mismatch values are $\delta=1.25, 2, 4$, and $7$, respectively. Dashed sections indicate unstable or oscillatory solutions. (b) A generalized $CE$ map showing the relationship between maximum $CE$, normalized pump power, $X$, and $\delta$. The white dashed line is expressed as $X=8\delta$ and closely follows the contour of highest maximum $CE$. (c) $CE$ (gray) and effective pump-resonator detuning, $\Delta_{\rm{p}}$ (green), versus $\alpha$. (d) $\Delta_{\rm{p}}$ versus $\delta$ at values of $\alpha$ that maximize $CE$ for $P_{\rm{in}}=15$ mW (red, $X\approx 5$) and $P_{\rm{in}}=25$ mW (orange, $X\approx 8$).} 
    \label{fig:etwm}
\end{figure*}

Figure \ref{fig:intro}c introduces some key concepts and illustrates differences between the TMA and the multi-mode LLE (here, 'multi-mode' refers to the inclusion of many longitudinal modes from the same spatial mode family). The main difference is that a multi-mode model accounts for many more modes that can be coupled together by the Kerr nonlinearity. In contrast, a TMA only considers nonlinear interactions (e.g., FWM) between the pump, signal, and idler modes. Importantly, both models account for imperfect frequency matching. In general, phase-matched mode pairs (i.e., with azimuthal numbers $\pm \mu$) are not frequency matched -- the associated FWM process does not conserve energy. Imperfect frequency matching is often quantified by the frequency mismatch parameter
\begin{equation}{\label{eq:mis}}
\delta_{\rm{\mu}}=\frac{\omega_{\mu}+\omega_{-\mu}-2\omega_{\rm{0}}}{\kappa(0)}. \end{equation} Throughout, we use $\delta_{\rm{\mu}}$ to refer to the dispersive mismatch spectrum and $\delta$ to refer to the value of $\delta_{\rm{\mu(-\mu)}}$ at the targeted signal (idler) mode. A degenerate FWM process only conserves energy and momentum if $\delta$ is compensated by nonlinear frequency shifts of $\omega_{\rm{p(s,i)}}$, which we denote as $N_{\rm{p (s, i)}}$ for  the pump (signal, idler) mode. Nonlinear shifts arise from self- and cross-phase modulation (SPM and XPM, respectively) and are related to the intraresonator intensity spectrum. Indeed, an occupied resonator necessarily implies $N_{p(s, i)}<0$; therefore, $\delta=0$ is not conducive to parametric oscillation.  %Importantly, this means resonators should not be designed for conventional phase matching (i.e. $\delta=0$), but should rather account for nonlinear shifts induced by the $\mu$OPO state.  

Using a TMA, we investigate how changes in $\delta$ impact the $\mu$OPO, and then we introduce other modes into our simulations. To ensure the consistency of our numerical methods, we use the split-step Fourier method to simulate Eq. \ref{eq:LLE} for both the TMA and multi-mode LLE. Our primary goal is to understand the efficiency with which pump photons are converted into signal or idler photons. Accordingly, we define the conversion efficiency in units of photon flux as
\begin{equation}{\label{eq:CE}}
CE=\frac{\omega_{\rm{p}}}{\omega_{\rm{s (i)}}}\frac{P_{\rm{s (i)}}}{P_{\rm{in}}}, 
\end{equation}
where $\omega_{\rm{s (i)}}$ is the signal (idler) frequency and $P_{\rm{s (i)}}$ is the signal (idler) output power in the access waveguide. Importantly, the underlying symmetry of the FWM process implies that the signal and idler fields have equal $CE$ values and that $N_{\rm{s}}=N_{\rm{i}}$ for a critically-coupled resonator. Throughout, we present $CE$ for the signal; the same results apply to the idler. 

In Fig. \ref{fig:etwm}, we present simulation results obtained using a TMA. In a single simulation, $P_{\rm{in}}$ and $\delta$ are held fixed; we vary $\alpha$ to simulate a $\omega_{\rm{p}}$ scan across resonance from blue to red detuning, as shown in Fig. \ref{fig:etwm}a. During the simulation, we record various data, including the $CE$, $N_{\rm{p(s,i)}}$, and optical spectra (see Supplemental Material for details). In Fig. \ref{fig:etwm}a, each panel corresponds to $P_{\rm{in}}=20$ mW, but $\delta$ is increased from top to bottom. Our results indicate a highest obtainable $CE$ of $12.5$~\% for critically-coupled resonators, in agreement with Ref. \cite{sayson2019octave}. Additionally, each panel depicts the effective signal detuning, defined as $\Delta_{\rm{s}}=\delta+\frac{\alpha}{2}-N_{\rm{s}}$. Notably, when $CE>0$, $\Delta_{\rm{s}} \approx 0$, which indicates that dispersion is perfectly compensated by detuning and nonlinearity in the $\mu$OPO state. 

To fully characterize the relationship between $CE$, $P_{\rm{in}}$, and $\delta$, we construct the universal $CE$ map shown in Fig. \ref{fig:etwm}b. The parameter space is defined by $P_{\rm{in}}$ and $\delta$; in Fig. \ref{fig:etwm}b, we normalize $P_{\rm{in}}$ as
\begin{equation}{\label{eq:norm}}
X=\frac{P_{\rm{in}}}{P_{\rm{th}}},
\end{equation}
where $P_{\rm{th}}=\frac{\hbar\omega_{\rm{0}}\kappa^3(0)}{8g_{\rm{0}}\kappa_{\rm{c}}(0)}$ is the oscillation threshold power \cite{kippenberg2004kerr}. Each data pixel in the $CE$ map represents the maximum $CE$ value taken from the corresponding LLE simulation, as indicated by the dashed line connecting Figs. \ref{fig:etwm}a and \ref{fig:etwm}b. The $CE$ map has a few notable features. First, we do not observe parametric oscillation for any values of $P_{\rm{in}}$ when $\delta\leq0$. This indicates that nonlinear frequency shifts of $\omega_{\rm{\mu}}$ inhibit frequency matching even when $X\approx 1$. Second, $CE$ contours follow clear trends through the parameter space. In particular, to maintain $CE$ at larger values of $X$, $\delta$ must be increased, apparently to compensate for larger $N_{s(p,i)}$. Remarkably, we can derive an analytical expression for the contour of highest $CE$. We provide the derivation in the Supplemental Material; here, we present the final result, $X=8\delta$, and indicate it with the white dashed line in Fig. \ref{fig:etwm}b. Clearly, to maximize $CE$ for a given $P_{\rm{in}}$, the microresonator dispersion (i.e., $\delta$) should be designed appropriately.
\begin{figure}[t]
    \centering
    \includegraphics[width=245 pt]{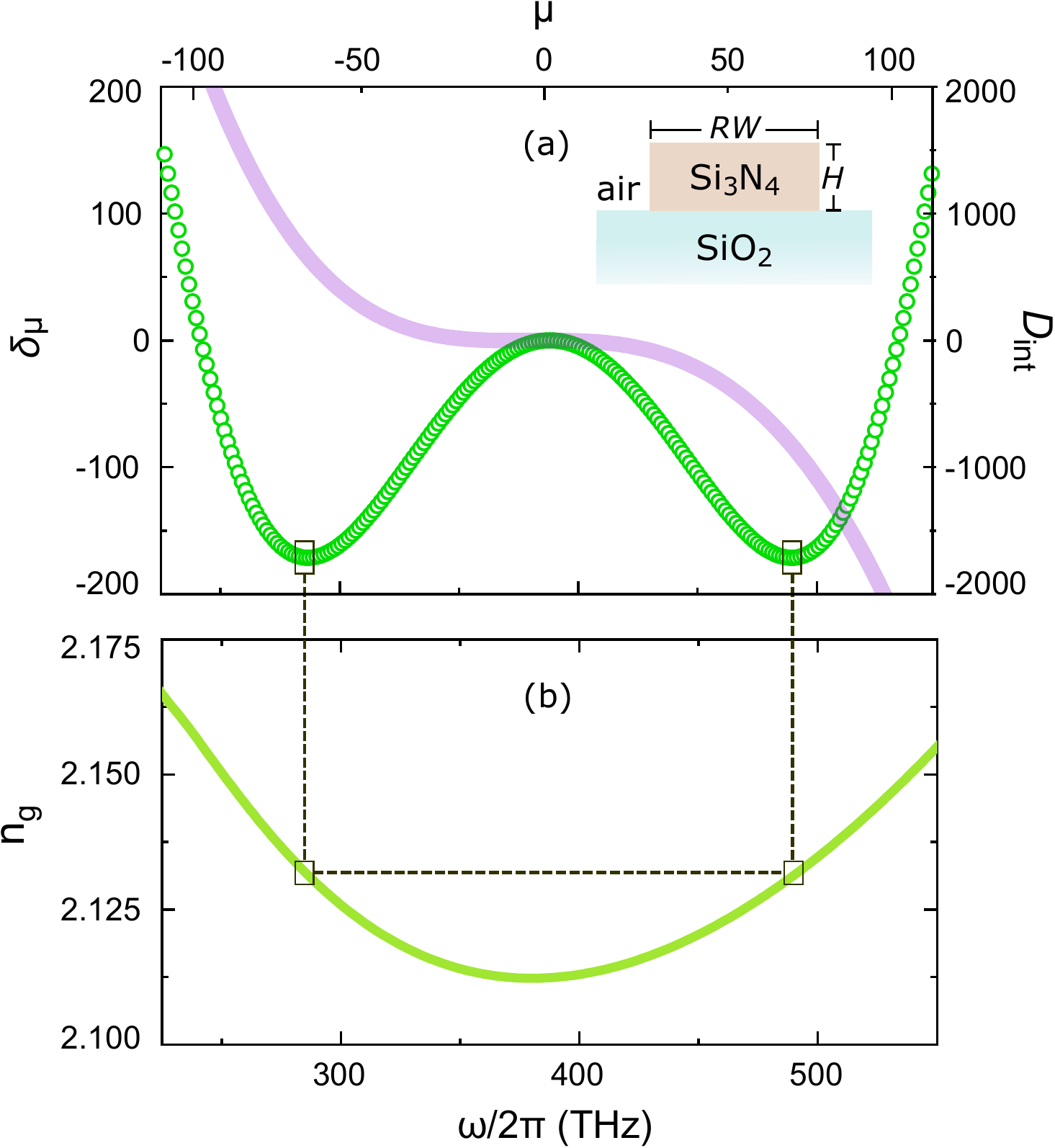}
    \caption{Depictions of microresonator dispersion. (a) $\delta_{\rm{\mu}}$ (green circles), and integrated dispersion, $D_{\rm{int}}$ (purple stripe), for the TE1 mode family calculated using finite-element method eigenfrequency simulations. The microresonator has ring radius $RR=15$ $\mu$m, ring width $RW=800$ nm, height $H=600$ nm, and pump frequency $\omega_{\rm{p}}=2\pi\times 388$ THz. Inset: Illustration of the test device cross section. (b) Group velocity refractive index, $n_{\rm{g}}$, versus $\omega/2\pi$ for the same device as in (a). The dashed lines illustrate how a turning point in $\delta_{\rm{\mu}}$ occurs when $n_{\rm{g}}(-\mu)=n_{\rm{g}}(\mu)$.}
    \label{fig:disp}
\end{figure}

In Fig. \ref{fig:etwm}a, it is perhaps remarkable that $CE$ values tend to increase monotonically until an abrupt cutoff. To elucidate this observation, we monitor the effective pump detuning, $\Delta_{\rm{p}}=\frac{\alpha}{2}-N_{\rm{p}}$, during our simulations and present a sample result in Fig. \ref{fig:etwm}c. When $CE$ is greatest, $\Delta_{\rm{p}}$ reaches a minimum value near zero. Physically, this condition implies that $\omega_{\rm{p}}$ is nearly resonant; intuitively, this is necessary to maximize the dropped power and, in turn, the nonlinear gain. The cutoff results from Kerr bistability; i.e., when further increases in $\alpha$ are no longer compensated by $N_{\rm{p}}$, the intraresonator field abruptly transitions to the CW state. To characterize the universality of this feature, we record $\Delta_{\rm{p}}$ (evaluated for the highest $CE$) versus $\delta$ for two different powers, as shown in Fig. \ref{fig:etwm}d. Apparently, realizing high $CE$ requires $\Delta_{\rm{p}}\approx 0$.

As a first step towards transitioning from a TMA to a multi-mode LLE, we briefly discuss the microresonator dispersion and its different representations. In Fig. \ref{fig:disp}a, we present $\delta_{\rm{\mu}}$ and $D_{\rm{int}}$ spectra for the fundamental transverse electric (TE) mode family, labeled TE1, of a SiN microring resonator (hereafter referred to as our test device) with ring radius $RR=15$ $\mu$m, ring width $RW=800$ nm, and height $H=600$ nm. We extract $\delta_{\rm{\mu}}$ and $D_{\rm{int}}$ from the mode spectrum, $\omega_{\rm{\mu}}$, that we calculate from finite-element method eigenfrequency simulations \footnote{Finite element simulations performed using Comsol Multiphysics, which is identified here to foster understanding, without implying recommendation or endorsement by NIST.}. It is easy to show that $\delta_{\rm{\mu}}=2\times \mathcal{S}(D_{\rm{int}}(\mu))$, where $\mathcal{S}(D_{\rm{int}}(\mu))$ indicates the symmetric part of $D_{\rm{int}}$; i.e., the even orders of $D_{\rm{int}}$ when expressed as a Taylor expansion around $\mu=0$. We have assessed that analyzing $\delta_{\rm{\mu}}$ is sufficient to understand the $\mu$OPO dynamics, which are not sensitive to odd orders of $D_{\rm{int}}$ \cite{sayson2019octave}. This is a notable departure from other nonlinear microresonator eigenstates, e.g., dissipative Kerr solitons. 

There are two defining features of the $\delta_{\rm{\mu}}$ spectrum shown in Fig. \ref{fig:disp}a. First, its negative curvature around $\mu=0$ indicates normal dispersion, which is required to suppress comb formation. Second, to overcome the normal dispersion and achieve frequency matching, the $D_{\rm{int}}$ expansion must contain higher order (even) terms such that $\delta_{\rm{\mu}}$ turns and becomes positive. Physically, this means that an anomalous-to-normal dispersion transition is necessary. Indeed, the dashed lines connecting Figs. \ref{fig:disp}a and \ref{fig:disp}b illustrate how $n_{\rm{g}}(\mu)=n_{\rm{g}}(-\mu)$ corresponds to a $\delta_{\rm{\mu}}$ turning point, where $n_{\rm{g}}(\mu)$ is the dispersive group velocity refractive index. %We stress that the properties of $\delta_{\rm{\mu}}$ identified above are broadly applicable to $\mu$OPO.   

\begin{figure}[ht]
    \centering
    \includegraphics[width=245 pt]{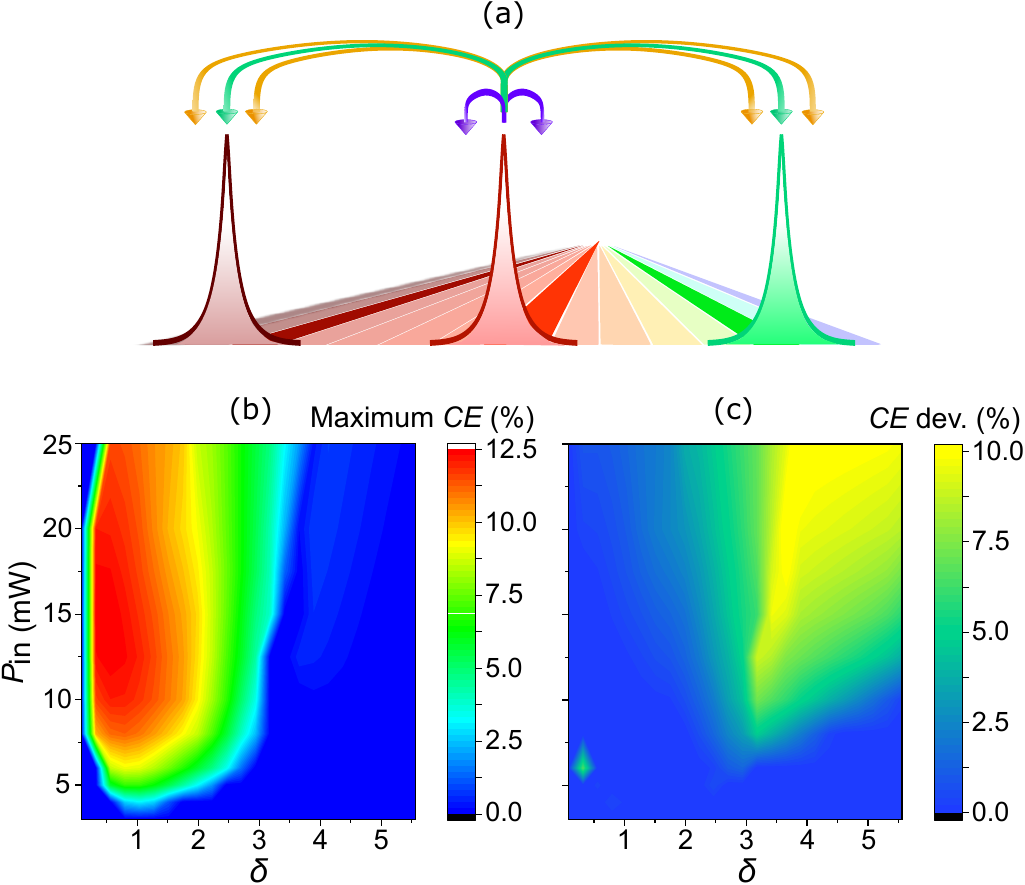}
    \caption{Transition from a TMA to multi-mode LLE simulations. (a) Depiction of the four wave mixing (FWM) processes studied in this work. Mode competition (orange arrows) suppresses FWM to the target signal and idler mode pair (with mode numbers $\pm \mu$) in favor of spectrally-adjacent mode pairs (with mode numbers $\pm (\mu \pm 1)$), and modulation instability induced by cross-phase modulation (XPM-MI, purple arrows) reroutes energy in the pump mode to its spectrally-nearest neighbors. (b) $CE$ map for mode $\mu=46$; the resonator parameters are $RR=15 \mu$m, $RW=800$ nm, $H=600$ nm, and $\omega_{\rm{p}}=2\pi\times 382$ THz. (c) Difference in $CE$ between simulations based on a TMA and the $CE$ map in part (b).}
    \label{fig:int}
\end{figure}

Next, we preview the differences between $CE$ maps calculated from multi-mode and TMA models. We observe two FWM processes, depicted in Fig. \ref{fig:int}a, that require the multi-mode model. In one case, we observe mode competition between mode pairs with consecutive $\mu$ values. In the second case, we observe modulation instability (MI) in the normally-dispersive spectral region around $\omega_{\rm{p}}$, and we explain it as arising from XPM between the pump, signal, and idler fields. We mostly constrain our multi-mode LLE simulations to $P_{\rm{in}}\leq 25$ mW ($X\leq 8$); at higher values of $P_{\rm{in}}$, the $\mu$OPO dynamics become more complex.  
\begin{figure*}
    \centering
    \includegraphics[width=500 pt]{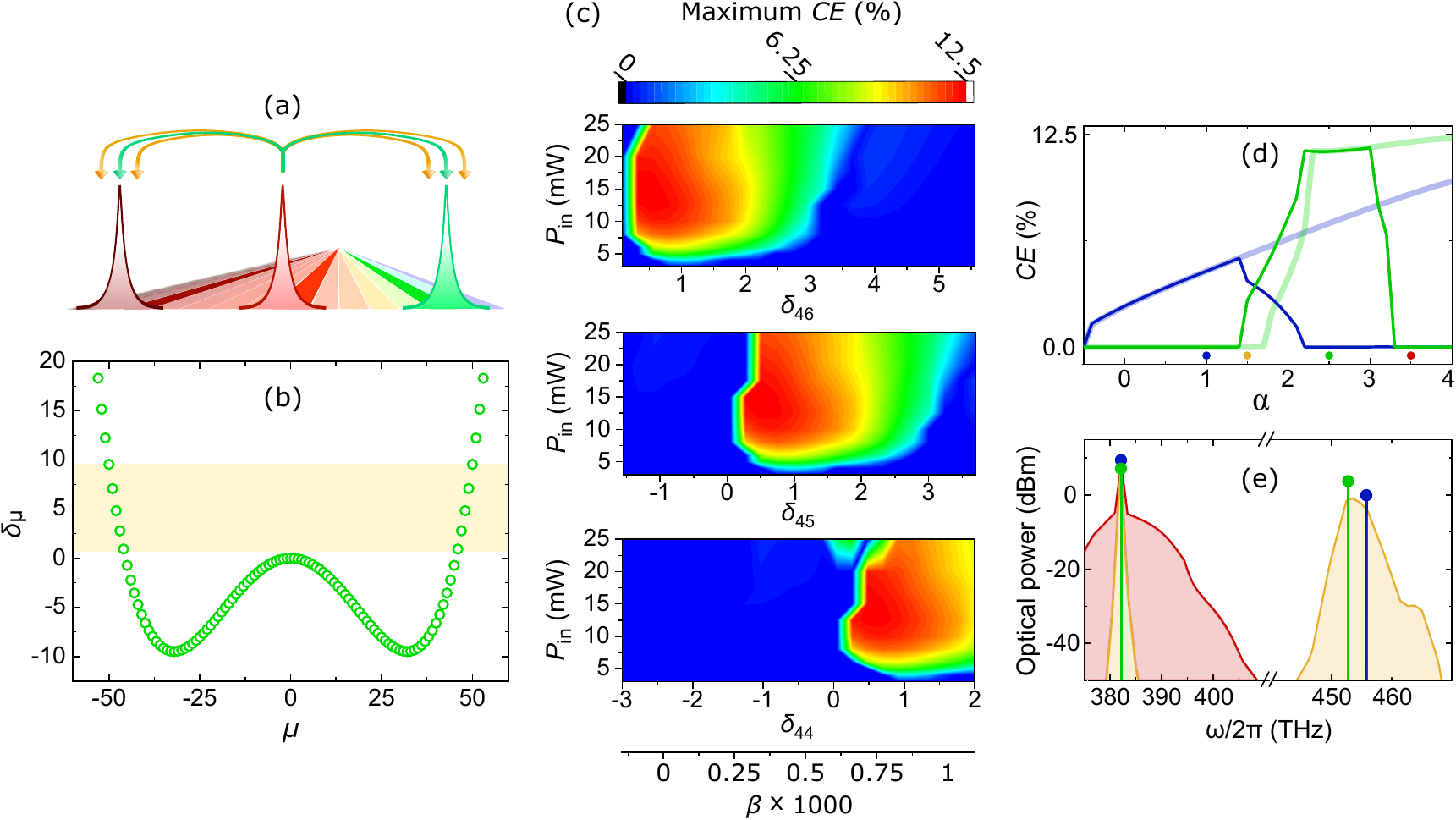}
    \caption{Mode competition and switching. (a) Illustration of mode competition in a $\mu$OPO. Oscillations on the target signal and idler mode pair (with mode numbers $\pm \mu$) are suppressed in favor of spectrally-adjacent mode pairs (with mode numbers $\pm (\mu \pm 1)$). (b) $\delta_{\rm{\mu}}$ for the test device with $\omega_{\rm{p}}=2\pi\times 382$ THz. The pale orange stripe overlaps modes that are predicted to oscillate when $P_{\rm{in}}=20$ mW. (c) $CE$ maps for modes $\mu=46$, $\mu=45$, and $\mu=44$. The maps indicate a mode competition in which the mode pair with smallest $\delta$ is favored for oscillation. (d) $CE$ versus $\alpha$ for modes $\mu=46$ (blue) and $\mu=45$ (green), where $P_{\rm{in}}=20$ mW and $\beta=5\times10^{-4}$. The pale stripes come from a TMA. (e) Optical spectra that correspond to different values of $\alpha$ in (d). The blue, orange, green, and red spectra correspond to $\alpha=1, 1.5, 2.5$, and $3.5$, respectively.}
    \label{fig:mcas}
\end{figure*}

Figure \ref{fig:int}b shows a $CE$ map for our test device pumped near $\omega_{\rm{p}}=2\pi\times 382$ THz. To make a straightforward comparison between these data and Fig. \ref{fig:etwm}b, we calculate the difference in $CE$ between the two $CE$ maps, as shown in Fig. \ref{fig:int}c. In general, XPM-induced MI (XPM-MI) explains $CE$ differences for small $\delta$, while mode competition is responsible for the abrupt cutoff in $CE$ (marked by the sharp transition to zero $CE$ in Fig. \ref{fig:int}b or the bold yellow stripe in Fig. \ref{fig:int}c) that indicates a different mode pair is oscillating. In the following sections, we explore mode competition and XPM-MI in detail. 

\section{Mode competition and switching}{\label{sec:mcas}}

Mode competition is ubiquitous in laser systems with multi-mode resonators \cite{narducci1986mode,gong2007numerical}. In general, mode competition occurs when several resonator modes experience amplification simultaneously; hence, modes compete for gain and become coupled. In this section, we present the results of multi-mode LLE simulations in which several mode pairs are simultaneously nearly frequency matched. We use our findings to establish general principles for mode competition that will inform future microresonator designs.

In Fig.~\ref{fig:mcas}, we present simulation results for our test device pumped near $\omega_{\rm{p}}=2\pi \times 382$ THz. The FWM process related to mode competition is illustrated in Fig.~\ref{fig:mcas}a. Modes that are spectrally adjacent to the targeted signal and idler pair may be nearly frequency matched; hence, light in these modes may become amplified through FWM. To explore this phenomenon, we consider the $\delta_{\rm{\mu}}$ spectrum shown in Fig. \ref{fig:mcas}b. According to the TMA, any mode pairs with $\delta_{\rm{\mu}}>0$ may oscillate, provided $P_{\rm{in}}$ is large enough. In Fig. \ref{fig:mcas}b, the data points covered by the pale gold stripe indicate mode pairs that would oscillate when $P_{\rm{in}}=20$ mW, if they (along with the pump mode) were the only modes in the system (i.e., in a TMA). Hence, we endeavour to understand how a mode pair (or pairs) is chosen for oscillation over its spectral neighbors. To study mode competition in our test device, we perform multi-mode LLE simulations, from which we construct $CE$ maps for the modes corresponding to $\mu=46$, $\mu=45$, and $\mu=44$, as shown in Fig. \ref{fig:mcas}c. To vary $\delta_{\rm{\mu}}$ (which is fixed for a given geometry), we apply a quadratic dispersion to $\omega_{\rm{\mu}}$, such that $\omega_{\rm{\mu}} \rightarrow \omega_{\rm{\mu}}+\beta \kappa(0)\mu^2$, where $\beta$ quantifies the added dispersion. We choose this approach in order to maintain the overall shape of $\delta_{\rm{\mu}}$. Note that $\delta_{\rm{45}}$ and $\delta_{\rm{44}}$ are both negative in Fig. \ref{fig:mcas}b, but they will become positive as $\beta$ is increased.  
\begin{figure*}[ht]
    \centering
    \includegraphics[width=500 pt]{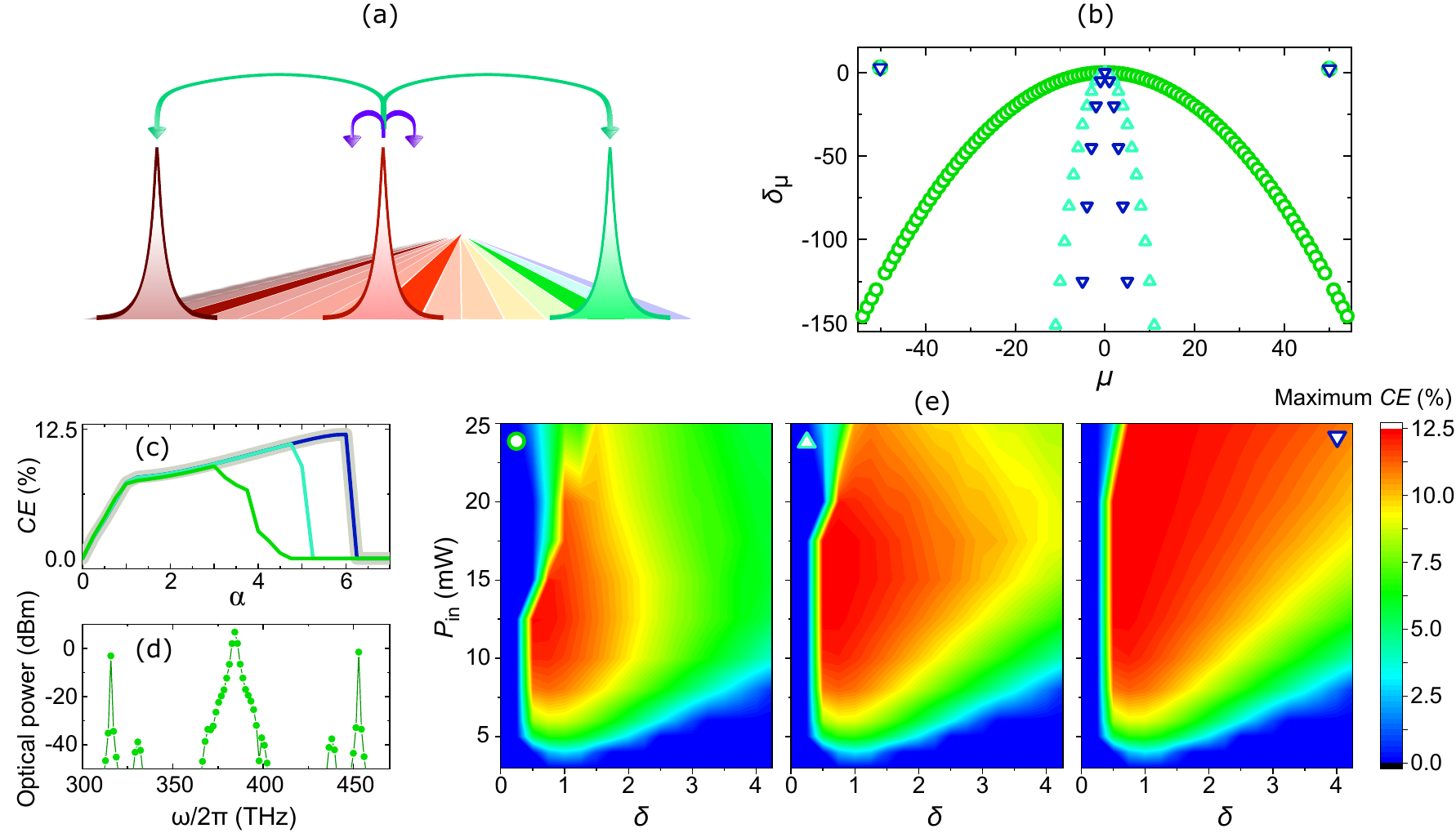}
    \caption{Characterization of XPM-MI in microresonators with normal dispersion. (a) Illustration of the XPM-MI FWM process (purple arrows), in which XPM between the pump, signal, and idler fields induces MI that distributes pump energy into spectrally-nearby sidebands. (b) $\delta_{\rm{\mu}}$ for $\beta=0.0625$ (green circles), $\beta=1.25$ (cyan upward-pointing triangles), and $\beta=5$ (blue downward-pointing triangles). One mode pair ($\mu=\pm 50$) is frequency shifted to be frequency matched to the pump mode. (c) $CE$ versus $\alpha$ for the $\delta_{\rm{\mu}}$ spectra in (b), with $P_{\rm{in}}=20$ mW and $\delta=3$. (d) Optical spectrum associated with part (c), with $\alpha=4$ and $\beta=0.0625$. (e) $CE$ maps for microresonators with the $\delta_{\rm{\mu}}$ spectra in (b), ordered from least to greatest $\beta$.}
    \label{fig:mic}
\end{figure*}

The $CE$ maps in Fig.~\ref{fig:mcas}c present clear evidence for mode competition, and we identify three notable features in them. First, the $CE$ map for mode $\mu=46$ resembles that of the TMA for small $\delta_{\rm{46}}$. Second, as $\beta$ is increased, the maximum $CE$ for mode $\mu=46$ declines sharply, and this coincides with $\delta_{\rm{45}}>0$ and oscillation on mode $\mu=45$. This pattern repeats as $\beta$ is increased further -- the oscillating mode changes from $\mu=45$ to $\mu=44$, and so on. Hence, we assess that the mode pair with smallest positive $\delta$ is favored for oscillation. Finally, there are small regions of parameter space where multiple mode pairs seem to oscillate simultaneously -- we explore this phenomenon in Figs. \ref{fig:mcas}d and \ref{fig:mcas}e. 

Figure \ref{fig:mcas}d shows $CE$ for modes $\mu=46$ and $\mu=45$ during a $\omega_{\rm{p}}$ scan (i.e., varying $\alpha$) with $\beta=5\times 10^{-4}$ ($\delta_{\rm{46}}\approx3$) and $P_{\rm{in}}=20$ mW. During the scan, mode $\mu=46$ begins to oscillate first; its $CE$ closely follows predictions made using a TMA (pale blue stripe) and reaches a maximum when $\alpha \approx 1.3$. In this regime, the $\mu$OPO spectrum is tri-modal, as depicted by the blue spectrum in Fig. \ref{fig:mcas}e (the spectral bandwidth in Fig. \ref{fig:mcas}e only spans $\omega_{\rm{p}}$ and $\omega_{\rm{s}}$; we have confirmed the spectrum is symmetric around $\omega_{\rm{p}}$). Beyond $\alpha \approx 1.3$, mode $\mu=45$ begins to oscillate, and $CE$ for mode $\mu=46$ deviates from its TMA counterpart. Between $\alpha \approx 1.3$ and $\alpha \approx 2.2$, both modes oscillate simultaneously, but their respective $CE$ values are not predicted by a TMA. Moreover, in this region the $\mu$OPO spectrum is not tri-modal; rather, it is strongly multi-moded with intensity peaks occurring near $\omega_{\rm{p}}$, $\omega_{\rm{s}}$, and $\omega_{\rm{i}}$, as shown by the orange spectrum in Fig. \ref{fig:mcas}e. Beyond $\alpha \approx 2.2$, $CE$ for mode $\mu=46$ is zero, and $CE$ for mode $\mu=45$ follows its TMA counterpart. We term this phenomenon, wherein $CE$ values for different mode pairs conform to their TMA counterparts for different portions of a $\omega_{\rm{p}}$ scan, mode switching. Finally, at $\alpha\geq 3$, the $\mu$OPO decays in favor of MI. Remarkably, the MI state is supported in the normally-dispersive region around $\mu=0$ and without strong signal or idler fields. The MI spectrum is shown by the shaded red curve in Fig. \ref{fig:mcas}e. In the next section, we explore MI and its impact on $CE$.

\section{XPM-MI: Characterization and Theory}{\label{sec:mict}}

It is clear from Fig. \ref{fig:mcas} that mode competition is not the only process that differentiates the multi-mode LLE and TMA. Specifically, Figs. \ref{fig:mcas}d and \ref{fig:mcas}e establish that MI can suppress or extinguish the signal and idler fields, and this occurs in regions of parameter space where a TMA predicts efficient parametric oscillation. In this section, we isolate the MI process in our simulations by considering a special dispersion profile that eliminates mode competition, and we develop a theory for MI as arising from XPM between the pump, signal, and idler fields. We term this process XPM-MI to link it to previous investigations \cite{agrawal1987modulation}.    

The XPM-MI FWM process is illustrated by the purple arrows in Fig. \ref{fig:mic}a. Energy from the pump laser is distributed to sidebands in a narrow spectral bandwidth around $\omega_{\rm{p}}$. If we use a multi-mode LLE to simulate our test device, the effects of mode competition can obfuscate the impact of XPM-MI. Therefore, we contrive the heuristically useful $\delta_{\rm{\mu}}$ spectra depicted in Fig. \ref{fig:mic}b. Here, we define $\delta_{\rm{\mu}}=-2\beta \mu^2$ for all $\mu \neq \pm 50$. The modes $\mu= \pm 50$ are designated for parametric oscillation and assigned a frequency mismatch value $\delta$ that can be manipulated apart from $\beta$. 
\begin{figure}[ht]
    \centering
    \includegraphics[width=245 pt]{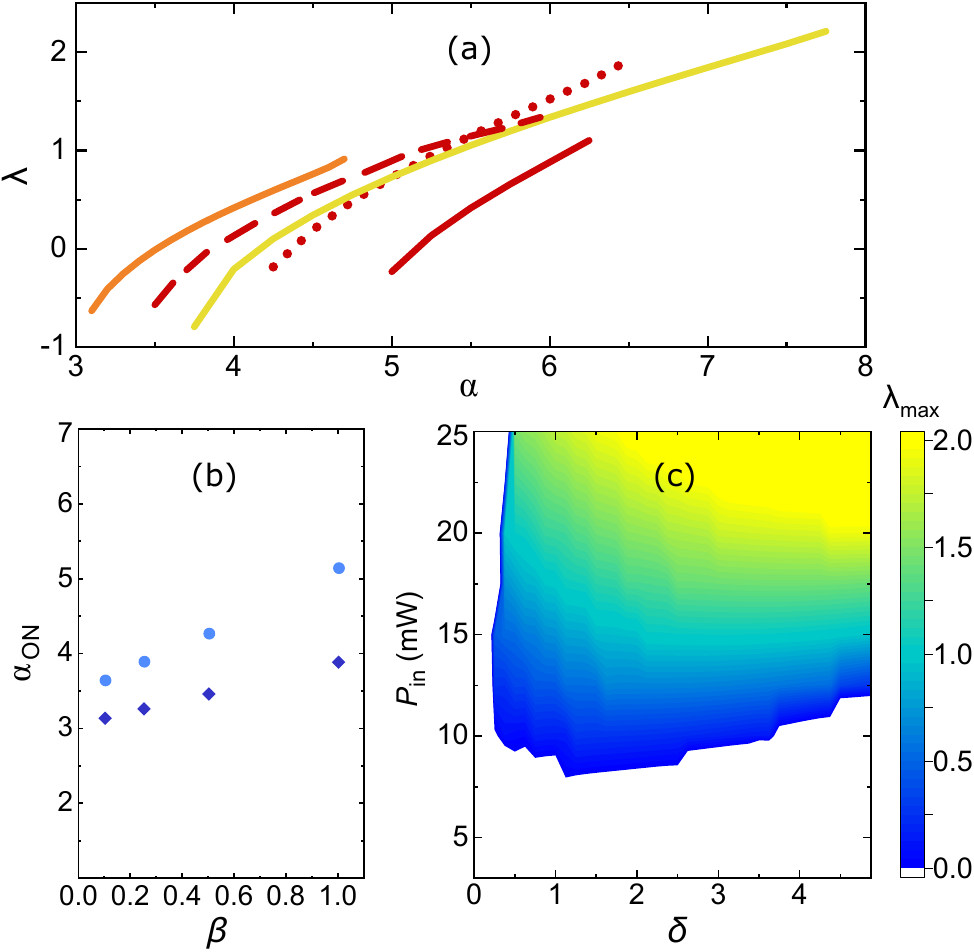}
    \caption{Theory of XPM-MI. (a) MI gain parameter, $\lambda$, versus $\alpha$ for $\mu=1$. Orange and gold curves correspond to $P_{\rm{in}}=15$ mW, $\delta=1.5$, $\beta=0.25$, and $P_{\rm{in}}=25$ mW, $\delta=2.5$, $\beta=0.25$, respectively. The red curves all use $P_{\rm{in}}=20$ mW, with bold, dotted, and dashed curves corresponding to $\delta=2.5$ and $\beta=1$, $\delta=5$ and $\beta=0.25$, and $\delta=2$ and $\beta=0.25$, respectively. (b) The value of $\alpha$ at which MI first appears, $\alpha_{\rm{ON}}$, versus $\beta$ for $P_{\rm{in}}=20$ mW and $\delta=2.5$. Light blue circles indicate theoretical values using Eq. \ref{eq:gain}, while dark blue diamonds mark values predicted by LLE simulations. (c) $\lambda$ map for $\mu=1$ and $\beta=0.25$.}
    \label{fig:mit}
\end{figure}

In Fig.~\ref{fig:mic}c, we present LLE simulations of $CE$ versus $\alpha$ for microresonators with the $\delta_{\rm{\mu}}$ spectra in Fig.~\ref{fig:mic}b; these data overlay the corresponding simulation using a TMA (pale gray stripe). They are representative of the entire ($P_{\rm{in}}, \delta)$ parameter space, and they are noteworthy for two reasons. First, in all cases $CE$ follows the gray stripe until $\alpha \geq \alpha_{\rm{ON}}$, where $\alpha_{\rm{ON}}$ corresponds to the onset of XPM-MI. An example XPM-MI spectrum is shown in Fig. \ref{fig:mic}d. Unlike the XPM-MI spectrum shown in Fig. \ref{fig:mcas}e, here the XPM-MI and $\mu$OPO states co-exist, albeit with suppressed $CE$. Second, $\alpha_{\rm{ON}}$ increases with increasing $\beta$. As a result, $CE$ fully converges to its TMA counterpart in the high-$\beta$ limit. To investigate this convergence, we construct the $CE$ maps for different values of $\beta$, and we present the results in Fig.~\ref{fig:mic}e. We observe steady convergence to the TMA $CE$ map as $\beta$ is increased. Clearly, it is crucial to realize strong normal dispersion near $\omega_{\rm{p}}$ to observe efficient parametric oscillation when $X>>1$.

Next, we present an expression for the XPM-MI gain that we derive from a set of coupled mode equations (see Supplemental Material for details), and we analyze this expression to support our conclusion that XPM underlies the observed MI states. As noted in Ref. \cite{hansson2013dynamics}, MI may occur in normally-dispersive microresonators; however, it usually requires a hard excitation, i.e., the MI sidebands are not amplified from vacuum fluctuations. This fact explains how the MI state can sometimes persist after parametric oscillations decay. Still, in our simulations we have never observed MI emerge before parametric oscillations. Indeed, it was predicted in Ref. \cite{agrawal1987modulation} that XPM between two waves can make them modulationally unstable, even when one or both waves propagates in normally-dispersive media. Motivated by this study, we analyze a set of coupled mode equations that assume the pump, signal, and idler modes are occupied, and we derive the MI gain, $\lambda_{\rm{\mu}}$, as
\begin{equation}{\label{eq:gain}}
\lambda_{\rm{\mu}}=-1+\sqrt{I_{\rm{0}}^2+I_{\rm{s}}^2+2I_{\rm{0}}I_{\rm{s}}\rm{cos}(\Delta \phi)-\textit{k}_{\rm{\mu}}^2},
\end{equation}
where $I_{\rm{0}}$ and $I_{\rm{s}}$ are the photon numbers for the pump and signal modes, respectively, $\Delta \phi=2\phi_{\rm{p}}-\phi_{\rm{s}}-\phi_{\rm{i}}$ is the relative phase mismatch between the pump, signal, and idler fields, which are expressed as $\tilde{a}_{\rm{p (s, i)}}=\sqrt{I_{\rm{p (s, i)}}}e^{-i\phi_{\rm{p (s, i)}}}$, and $k_{\rm{\mu}}=2I_{\rm{0}}+4I_{\rm{s}}-\beta \mu^2-\alpha$. MI occurs when $\lambda_{\rm{\mu}}>0$, for any $\mu$. Note from these definitions that $\beta \mu^2$ is the frequency mismatch parameter for the MI sidebands, with mode numbers $\pm \mu$, and not the frequency mismatch for the $\mu$OPO mode pair, which we denote as $\delta$. In the limit $I_{\rm{s}} \rightarrow 0$, Eq. \ref{eq:gain} is equivalent to the MI gain expression derived in Ref. \cite{hansson2013dynamics}. Moreover, it is clear that $\lambda_{\rm{\mu}}$ is maximized when the pump, signal, and idler fields are in phase (i.e. when $\Delta \phi = 0$), which corresponds to the in-phase addition of two FWM processes: the degenerate FWM process $2\omega_{\rm{p}}\rightarrow \omega_{\rm{\mu}}+\omega_{\rm{-\mu}}$ and the non-degenerate process $\omega_{\rm{s}}+\omega_{\rm{i}}\rightarrow \omega_{\rm{\mu}}+\omega_{\rm{-\mu}}$.

To analyze XPM-MI using Eq.~\ref{eq:gain}, we perform LLE simulations, using a TMA, for various values of $P_{\rm{in}}$ and $\delta$ and extract values for $I_{\rm{0}}, I_{\rm{s}}$, and $\Delta \phi$ that we use to calculate $\lambda_{\rm{\mu}}$. When $\lambda_{\rm{\mu}}>0$, the mode pair with mode numbers $\pm\mu$ will be amplified and steal energy from the $\mu$OPO. In general, a large normal dispersion leads to the sideband pair with $\mu=\pm 1$ having the largest gain; therefore, in what follows we drop the $\mu$ subscript and define $\lambda$ as the MI gain for this pair. We have confirmed that $\lambda_{\rm{\mu}}<0$ whenever $I_{\rm{s}}=0$ and $\beta>0$; physically, this means that XPM between the pump, sigal, and idler fields is required to initiate MI for microresonators with normal dispersion. Hence, we term this process XPM-MI. Figure \ref{fig:mit}a presents $\lambda$ calculations for various values of $P_{\rm{in}}$, $\delta$, and $\beta$. They indicate that $\lambda$ grows with increasing $\alpha$, which is explained by the stronger XPM that results from more powerful $\mu$OPO sidebands at large $\alpha$. To compare our XPM-MI theory (i.e., Eq.~\ref{eq:gain}) to multi-mode LLE simulations, we calculate $\alpha_{\rm{ON}}$ in both cases, for different values of $\beta$. Overall, we observe good agreement, but Eq.~\ref{eq:gain} generally predicts higher values of $\alpha_{\rm{ON}}$ than the multi-mode LLE. A sample comparison of this type is presented in Fig. \ref{fig:mit}b.   
\begin{figure*}
    \centering
    \includegraphics[width=500 pt]{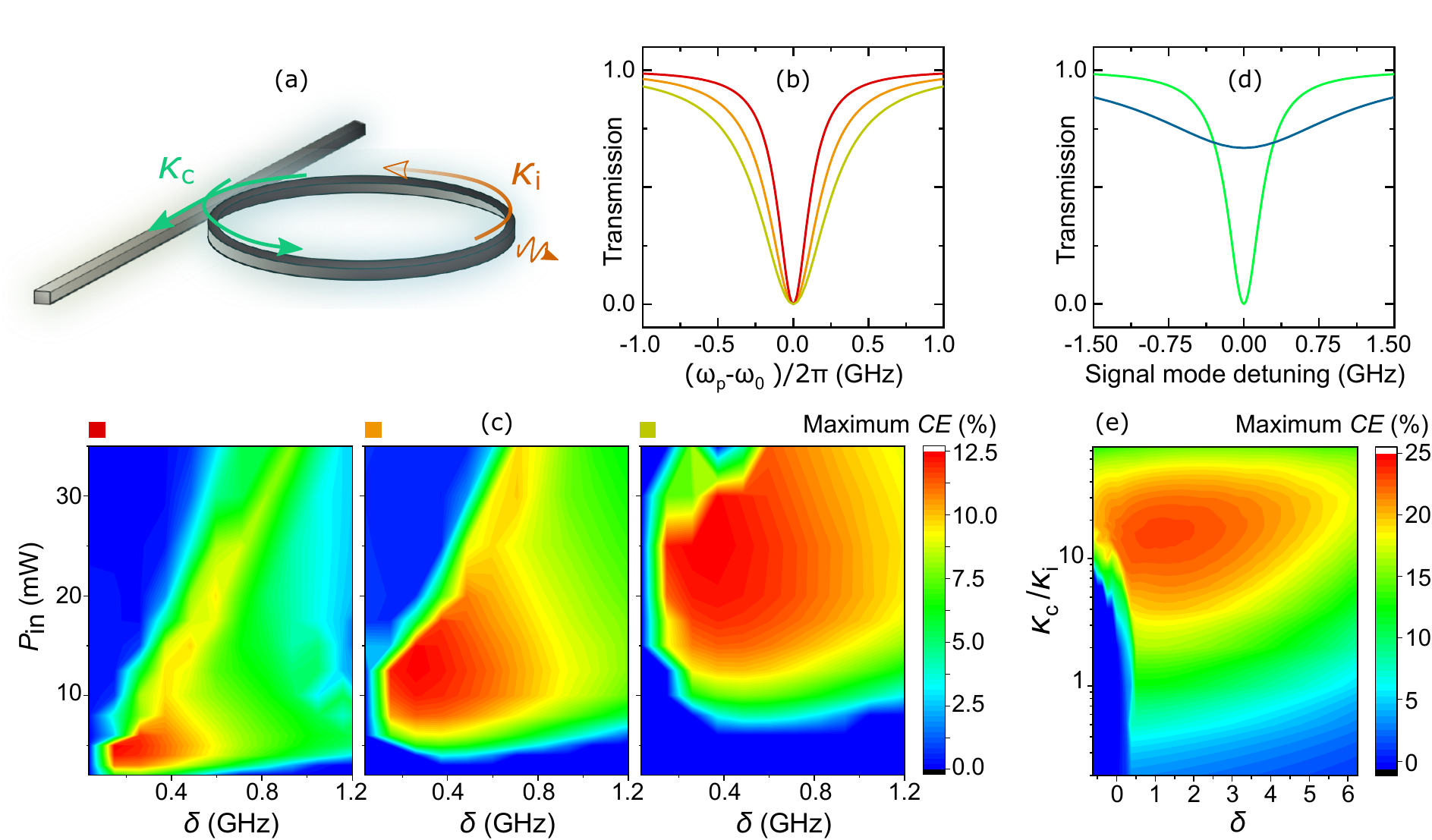}
    \caption{Strategies to achieve high $CE$ and increase power output from a $\mu$OPO. (a) Depiction of a microring resonator and its loss rates. (b) Simulated mode transmission vs $\omega_{\rm{p}}$ for $\kappa_{\rm{i}}=2\pi\times 125$ MHz (red), $\kappa_{\rm{i}}=2\pi\times 200$ MHz (orange), and $\kappa_{\rm{i}}=2\pi\times 275$ MHz (gold). In each case, $\kappa_{\rm{c}}=\kappa_{\rm{i}}$. (c) $CE$ maps corresponding to (b), for a resonator with $RR=15~\mu$m, $RW=800$ nm, $H=600$ nm, and $\omega_{\rm{p}}=2\pi\times 385$ THz. (d) Simulated mode transmissions for $\kappa_{\rm{c}}=\kappa_{\rm{i}}$ (green) and $\kappa_{\rm{c}}=10\times \kappa_{\rm{i}}$ (blue), where $\kappa_{\rm{i}}=2\pi\times 200$ MHz. (e) Maximum $CE$ versus $\kappa_{\rm{c}}/\kappa_{\rm{i}}$ and $\delta$, where $\kappa_{\rm{c}}$ is the signal mode coupling rate. The pump and idler modes are critically coupled.}
    \label{fig:prop}
\end{figure*}

To more comprehensively compare our theory with multi-mode LLE simulations, we calculate the maximum gain, $\lambda_{\rm{max}}$, for $\beta=0.25$ and different values of $P_{\rm{in}}$ and $\delta$. We use these data to construct the $\lambda_{\rm{max}}$ map presented in Fig. \ref{fig:mit}c. It accurately predicts the threshold $P_{\rm{in}}$ values for XPM-MI, and it has the overall trend that $\lambda_{\rm{max}}$ grows with increasing $P_{\rm{in}}$ and $\delta$. This trend is consistent with the $CE$ maps presented in Fig. \ref{fig:mic}e, which deviate from the TMA as $P_{\rm{in}}$ and $\delta$ are increased. Importantly, our observations may explain prior experimental results, which consistently report that increasing $P_{\rm{in}}$ leads to the formation of undesired sidebands \cite{lu2019milliwatt, tang2020widely, fujii2019octave}. Overall, our theory supports the hypothesis that XPM drives MI in a $\mu$OPO. Moreover, we can now form a concise description of parasitic FWM in $\mu$OPOs: Mode competition dictates which mode pair oscillates, while XPM-MI constrains $CE$. 

\section{Towards efficient and high-power $\mu$OPO}

Our results suggest several design rules for avoiding mode competition and mitigating XPM-MI through dispersion engineering. For example, Fig. \ref{fig:mic} demonstrates that larger normal dispersion around $\omega_{\rm{p}}$ will suppress XPM-MI. Moreover, increasing $\frac{d\delta_{\rm{\mu}}}{d\mu}$ will help prevent mode competitions. This can be accomplished by using smaller resonators or in resonators with greater dispersion. Still, in practice it is nontrivial to fabricate devices with ideal dispersion characteristics. Therefore, in this section we describe two strategies to increase $CE$ by tuning the microresonator loss rates, $\kappa_{\rm{c}}(\mu)$ and $\kappa_{\rm{i}}$. We assume that users of real-world $\mu$OPO devices will value high output power; therefore, we focus on ways to increase $CE$ for fixed $P_{\rm{in}}$.

Figure \ref{fig:prop}a depicts the coupling and intrinsic loss rates that are quantified by $\kappa_{\rm{c}}(\mu)$ and $\kappa_{\rm{i}}$, respectively. In practice, $\kappa_{\rm{c}}(\mu)$ is controllable by various design parameters, including the waveguide-resonator separation and waveguide-resonator coupling length, for example, in a `pulley' configuration ~\cite{moille2019broadband}. Moreover, $\kappa_{\rm{i}}$ can be reduced within some spectral bands by annealing \cite{spencer2014integrated,graziosi2018enhancement}; hence, one gains some control over $\kappa_{\rm{i}}$ by making a suitable choice for the annealing time or temperature (also, increasing $\kappa_{\rm{i}}$ can be realized through, for example, intentionally-introduced surface roughness). 

We consider two ways to increase $CE$. First, by increasing $P_{\rm{th}}$, one increases the $P_{\rm{in}}$ values for which XPM-MI occurs. To demonstrate this approach, we perform multi-mode LLE simulations of our test device for three different values of $\kappa$. In every simulation, $\kappa_{\rm{c}}(\mu)=\kappa_{\rm{i}}$ and $\omega_{\rm{p}}=2\pi \times 385$ THz. The simulated modal lineshapes for these resonators are shown in Fig. \ref{fig:prop}b, and the corresponding $CE$ maps are shown in Fig. \ref{fig:prop}c. In Fig. \ref{fig:prop}c, $\kappa$ is increased from the left-most panel to the right-most panel. As $\kappa$ is increased, the region of highest $CE$ is shifted towards higher values of $P_{\rm{in}}$. Moreover, this region is broadened along both axes, indicating that devices with larger $\kappa$ will have a greater tolerance for design errors in $\delta$. While increases in $\kappa$ prevent high $CE$ at low $P_{\rm{in}}$, the obtainable output power generated with high $P_{\rm{in}}$ has grown; for instance, with $P_{\rm{in}}=30$ mW, the maximum output power increases from $P_{\rm{sig}}\approx 2.25$ mW at $\kappa/2\pi=250$~MHz (Fig.~\ref{fig:prop}c, left-most panel) to $P_{\rm{sig}}\approx 3.75$ mW at $\kappa/2\pi=550$~MHz (Fig.~\ref{fig:prop}c, right-most panel).     

Finally, we consider the relationship between the signal mode coupling rate, $\kappa_{\rm{c}}$, and $CE$. This relationship was explored in Ref. \cite{sayson2019octave}, with the result that $CE$ may be as large as $25$\% (overcoupling the pump mode may further increase $CE$). Here, we reiterate this result and explore it within the $\mu$OPO parameter space. Figure \ref{fig:prop}d shows simulated signal mode lineshapes for two values of $\kappa_{\rm{c}}$, and Fig.~\ref{fig:prop}e shows the $CE$ map for $P_{\rm{in}}=20$ mW in a parameter space defined by the coupling ratio, $\kappa_{\rm{c}}/\kappa_{\rm{i}}$, and $\delta$. In our simulations, we use a TMA and keep the pump and idler modes critically coupled. We not only observe an increase in the highest obtainable $CE$ to nearly $25$\%, but we also observe that the region of highest $CE$ is broadened for increasing $\kappa_{\rm{c}}/\kappa_{\rm{i}}$. Broadening occurs until $\kappa_{\rm{c}}/\kappa_{\rm{i}}\approx 30$, at which point the advantages of overcoupling are overcome by the corresponding increases in $P_{\rm{th}}$.

\section{Discussion}

In conclusion, we have established a foundation of simulation results for $\mu$OPOs that moves beyond a TMA and will help guide experimental efforts to realize the high conversion efficiencies predicted by the simplified theory. We introduced a $CE$ map that encapsulates the $\mu$OPO solution space, and through multi-mode LLE simulations we reveal nonlinear dynamics not predictable from a TMA. In particular, we identified mode competition and demonstrated how it determines the oscillating mode pair. Meanwhile, the range of parameter space over which high $CE$ can be obtained is constrained by XPM-MI. Mode competition and XPM-MI both depend on the microresonator dispersion, $\delta_{\rm{\mu}}$, and suitable dispersion engineering may circumvent these processes. Still, optimizing $\delta_{\rm{\mu}}$ may be nontrivial in practice; therefore, we have proposed two strategies to increase $CE$ apart from dispersion engineering. Ultimately, suitable control of both resonator loss (including waveguide coupling) and dispersion makes it possible to tailor microresonator geometries for high $CE$ at a targeted pump power, that is, to produce a useful amount of output power. Such engineering will be crucial in the development of compact, coherent light sources that take advantage of the enormous wavelength flexibility inherent to $\chi^{(3)}$ OPOs. 

\begin{acknowledgments}
We thank Travis Briles and Edgar Perez for a careful reading of the paper. This project is funded by the DARPA LUMOS program.
\end{acknowledgments}

%\subsection{}
%\subsubsection{}

\clearpage

\onecolumngrid

\begin{center}
{\huge Supplemental Material}
\end{center}
\setcounter{section}{0}

\section{Calculations of $\mu$OPO variables}
In this section, we describe mathematical formulae that relate LLE variables to the data (e.g. $CE$ and optical spectra) presented in the main text. Simulations of the Lugiato-Lefever Equation (LLE) yield solutions for the complex intraresonator field, $a$; we denote the intraresonator field spectrum $\tilde{a}=\mathcal{F}(a)$, where $\mathcal{F}$ denotes the Fourier transform, and $a$ is normalized such that $|\tilde{a}|^2_{\omega=\omega_{\mu}}$ gives the number of intraresonator photons in the mode $\mu$. 

Table \ref{tab:1} lists the important variables, along with their physical descriptions and how they are calculated from $a$, $\tilde{a}$, and simulation parameters. Notably, the expressions for $\phi_{\rm{p}}$ and $N_{\rm{p}}$ also apply to the signal and idler modes when the expressions are evaluated at the appropriate value of $\omega$. We follow Ref. \cite{yu2021spontaneous} in our calculation of the nonlinear mode frequency shifts.

\begin{table}[h!]
\begin{center}
 \begin{tabular}{||c c c||} 
 \hline
 Variable & Description (units) & LLE Calculation \\ [0.5ex] 
 \hline\hline
 $P_{\rm{sig}}$ & Output signal power (W) & $\kappa_{\rm{c}} \hbar\omega_{\rm{s}}|\tilde{a}|^2_{\omega=\omega_{\rm{s}}}$ \\
 \hline
 $\phi_{\rm{p}}$ & Pump phase (rad) & $\angle \tilde{a}_{\omega=\omega_{\rm{0}}}$ \\
 \hline
 $N_{\rm{p}}$ & Pump mode nonlinear frequency shift ($\kappa(0)/2\pi$) & $\frac{1}{\kappa(0)}\left(\frac{\mathcal{F}(g_{\rm{0}}|a|^2a)}{\tilde{a}}\right)_{\omega=\omega_{\rm{0}}}$ \\
 \hline
 $P_{\rm{out}}$ & Pump laser output power ($W$) & $\hbar\omega_{\rm{p}}|\sqrt{\kappa_{\rm{c}}(0)}\tilde{a}_{\omega=\omega_{\rm{0}}}-\sqrt{\frac{P_{\rm{in}}}{\hbar\omega_{\rm{p}}}}|^2$ \\
 \hline
\end{tabular}
\end{center}
\caption{List and descriptions of important variables derived from the LLE.}
\label{tab:1}
\end{table}

\section{Derivation of highest maximum $CE$ contour}
Next, we seek to derive the result - stated in the main text - that the contour of highest maximum $CE$ in the TMA is given by $X=8\delta$. Our starting point is the following set of coupled equations \cite{sayson2019octave}: 
\begin{align}
    \frac{dA_{\rm{p}}}{dt} &=-(1+i\alpha)A_{\rm{p}}+i(|A_{\rm{p}}|^2+2|A_{\rm{s}}|^2+2|A_{\rm{i}}|^2)A_{\rm{p}}+2A_{\rm{p}}^*A_{\rm{s}}A_{\rm{i}}+\sqrt{X} \label{eq:S1} \\
    \frac{dA_{\rm{s}}}{dt} &=-(1+i\alpha+i\delta)A_{\rm{s}}+i(|A_{\rm{s}}|^2+2|A_{\rm{p}}|^2+2|A_{\rm{i}}|^2)A_{\rm{s}}+2A_{\rm{s}}^*A_{\rm{p}}A_{\rm{i}} \label{eq:S2} \\
    \frac{dA_{\rm{i}}}{dt} &=-(1+i\alpha+i\delta)A_{\rm{i}}+i(|A_{\rm{i}}|^2+2|A_{\rm{p}}|^2+2|A_{\rm{s}}|^2)A_{\rm{i}}+2A_{\rm{i}}^*A_{\rm{p}}A_{\rm{s}}, \label{eq:S3}
\end{align}
where $A_{\rm{p(s,i)}}=\sqrt{I_{\rm{p(s,i)}}}e^{-i\phi_{\rm{p(s,i)}}}$ denotes the pump (signal, idler) field, and $I_{\rm{p(s,i)}}=|A_{\rm{p(s,i)}}|^2$ is proportional to the pump (signal, idler) intraresonator photon number. If we substitute the field expressions into the above equations, we obtain six coupled equations for the real variables $I_{\rm{p(s, i)}}$ and $\phi_{\rm{p(s, i)}}$,
\begin{align}
\frac{dI_{\rm{p}}}{dt} &=-2I_{\rm{p}}-4I_{\rm{p}}\sqrt{I_{\rm{s}}I_{\rm{i}}}\textrm{sin}(\Delta\phi)+2\sqrt{I_{\rm{p}}X}\textrm{cos}(\phi_{\rm{p}})\\
\frac{dI_{\rm{s}}}{dt} &=-2I_{\rm{s}}+2I_{\rm{p}}\sqrt{I_{\rm{s}}I_{\rm{i}}} \textrm{sin}(\Delta \phi)\\
\frac{dI_{\rm{i}}}{dt} &=-2I_{\rm{i}}+2I_{\rm{p}}\sqrt{I_{\rm{s}}I_{\rm{i}}} \textrm{sin}(\Delta \phi)\\
\frac{d\phi_{\rm{p}}}{dt} &=I_{\rm{p}}+2I_{\rm{s}}+2I_{\rm{i}}+2\sqrt{I_{\rm{s}}I_{\rm{i}}}\textrm{cos}(\Delta \phi)-\alpha-\sqrt{\frac{X}{I_{\rm{p}}}}\textrm{sin}(\phi_{\rm{p}})\\
\frac{d\phi_{\rm{s}}}{dt} &=I_{\rm{s}}+2I_{\rm{p}}+2I_{\rm{i}}+I_{\rm{p}}\sqrt{\frac{I_{\rm{i}}}{I_{\rm{s}}}}\textrm{cos}(\Delta \phi)-\alpha-\delta\\
\frac{d\phi_{\rm{i}}}{dt} &=I_{\rm{i}}+2I_{\rm{p}}+2I_{\rm{s}}+I_{\rm{p}}\sqrt{\frac{I_{\rm{s}}}{I_{\rm{i}}}}\textrm{cos}(\Delta \phi)-\alpha-\delta,
\end{align}
where $\Delta \phi=2\phi_{\rm{p}}-\phi_{\rm{s}}-\phi_{\rm{i}}$. We next reduce this system to four equations by first noting that, from the symmetry of Eqs. \ref{eq:S2} and \ref{eq:S3}, $I_{\rm{s}}=I_{\rm{i}}$. Therefore, we define $M=I_{\rm{s}}=I_{\rm{i}}$. Moreover, momentum conservation implies that $\frac{d}{dt}(\phi_{\rm{s}}+\phi_{\rm{i}})=0$. Hence, for steady state conditions we obtain
\begin{align}
    I_{\rm{p}} &=-2I_{\rm{p}}M\textrm{sin}(\Delta \phi)+\sqrt{I_{\rm{p}}X}\textrm{cos}(\phi_{\rm{p}})\label{eq:s10}\\
    1 &=I_{\rm{p}}\textrm{sin}(\Delta \phi)\label{eq:s11}\\
    \sqrt{\frac{X}{I_{\rm{p}}}}\textrm{sin}(\phi_{\rm{p}}) &=I_{\rm{p}}+4M+2M\textrm{cos}(\Delta \phi)-\alpha \label{eq:s12}\\
    I_{\rm{p}}\textrm{cos}(\Delta \phi) &=\alpha-3M-2I_{\rm{p}}+\delta\label{eq:s13}.
\end{align}
Equations \ref{eq:s10}-\ref{eq:s13} are still sufficiently general to study the stationary solutions. However, we proceed to simplify them via two ansatzes associated with a maximally efficient $\mu$OPO. Specifically, we set $\phi_{\rm{p}}=0$, which corresponds to the pump laser being on resonance, and $M=\frac{I_{\rm{p}}}{2}$. The validity of these assumptions is confirmed by our numerical results, but they are also physically intuitive. For instance, because the FWM process that drives $\mu$OPO ($2\omega_{\rm{p}}\rightarrow \omega_{\rm{s}}+\omega_{\rm{i}}$) is reversible, one expects the intraresonator photons to be evenly distributed between the pump mode and sideband pair.  

After inserting our two ansatzes into Eqs. \ref{eq:s10}-\ref{eq:s13}, we combine Eqs. \ref{eq:s10} and \ref{eq:s11} to obtain $I_{\rm{p}}=X/4$. Then, Eqs. \ref{eq:s12} and \ref{eq:s13} are combined to obtain $\frac{I_{\rm{p}}}{2}=\delta$. Insertion of the former into the latter yields the desired result, $X=8\delta$.

\section{Derivation of XPM-MI gain}

Finally, in this section we derive Eq. 4 from the main text. We follow the classical procedure of linearizing a set of coupled equations around a steady-state solution and then allowing plane-wave perturbations to grow exponentially. We start from the equations of motion for the fields $A_{\rm{\mu}}$ and $A_{\rm{-\mu}}$ when the pump, signal, and idler modes are occupied,
\begin{align}
    \frac{dA_{\rm{\mu}}}{dt} &=i\beta\mu^2+i(2I_{\rm{p}}+2I_{\rm{s}}+2I_{\rm{i}}+2I_{\rm{-\mu}}+I_{\rm{\mu}})A_{\rm{\mu}}-(1+i\alpha)A_{\rm{\mu}}+iA_{\rm{\mu}}^*A_{\rm{p}}^2+iA_{\rm{\mu}}^*A_{\rm{s}}A_{\rm{i}}\\
    \frac{dA_{\rm{-\mu}}}{dt} &=i\beta\mu^2+i(2I_{\rm{p}}+2I_{\rm{s}}+2I_{\rm{i}}+I_{\rm{-\mu}}+2I_{\rm{\mu}})A_{\rm{-\mu}}-(1+i\alpha)A_{\rm{-\mu}}+iA_{\rm{-\mu}}^*A_{\rm{p}}^2+iA_{\rm{-\mu}}^*A_{\rm{s}}A_{\rm{i}},
\end{align}
where $\beta$ quantifies the dispersion. Next, we introduce the perturbation $\delta f_{\rm{\mu}}(t)$, so that the equations of motion for the perturbations, after simplifying, read
\begin{align}
      \frac{d\delta f_{\rm{\mu}}}{dt} &= (ik_{\rm{\mu}}-1)\delta f_{\rm{\mu}}+i(A_{\rm{p}}^2+A_{\rm{s}}A_{\rm{i}})\delta f_{\rm{\mu}}^*\\
      \frac{d\delta f_{\rm{-\mu}}^*}{dt} &= -(ik_{\rm{\mu}}+1)\delta f_{\rm{-\mu}}^*-i(A_{\rm{p}}^{*2}+A_{\rm{s}}^*A_{\rm{i}}^*)\delta f_{\rm{\mu}},
\end{align}
where $k_{\rm{\mu}}=\beta\mu^2+2I_{\rm{p}}+4I_{\rm{s}}-\alpha$, and we have again assumed $I_{\rm{s}}=I_{\rm{i}}$. If we set $\delta f_{\rm{\mu}}(t)=ae^{\lambda_{\rm{\mu}}t}$, then we obtain a set of linear homogenous equations with eigenvalue $\lambda_{\rm{\mu}}$. Solving the eigenvalue problem yields
\begin{equation}
    \lambda_{\rm{\mu}}=-1\pm\sqrt{I_{\rm{p}}^2+I_{\rm{s}}^2+2I_{\rm{p}}I_{\rm{s}}\textrm{cos}(\Delta \phi)-k_{\rm{\mu}}^2},
\end{equation}
which is the desired result. Notably, the original set of coupled equations are approximations in several ways. For instance, we only consider one sideband pair, but the full XPM-MI dynamics involves the collective interactions of multiple sideband pairs simultaneously. In addition, we neglect FWM processes that do not involve both MI sidebands (e.g. the interaction $A_{\rm{\mu}}^*A_{\rm{s}}^*A_{\rm{p}}A_{\rm{s+\mu}}$). These approximations may explain the small discrepancies between our theory and LLE simulations.

\bibliography{Bibliography}

\end{document}